%% file: paper.tex
\documentclass[letter,prd,aps,twocolumn,floatfix,superscriptaddress,nofootinbib]{revtex4}

\usepackage{amssymb,amsmath,graphicx}
\usepackage{hyperref}
\usepackage[usenames,dvipsnames]{color}
\usepackage{array}
\makeatletter
\renewcommand*{\p@section}{\S\,}
\renewcommand*{\p@subsection}{\S\,}

\makeatother

\input{aas_macros.tex}

\begin{document}

\date{\today}
\title{Scalar collapse in AdS with an OpenCL open source code}

\author{Steven L. Liebling}
\affiliation{Long Island University, Brookville, New York 11548, USA}
\author{Gaurav Khanna}
\affiliation{Department of Physics \& Center for Scientific Computing and Visualization Research,
      University of Massachusetts Dartmouth, North Dartmouth, MA 02747, USA}


\begin{abstract}
We study the spherically symmetric collapse of a scalar field in anti-de Sitter spacetime using a newly constructed, open-source code which parallelizes over heterogeneous architectures using the open standard OpenCL.
An open question for this scenario concerns how to tell, \textit{a priori},
whether some form of initial data will be stable or will instead develop under
the turbulent instability into a black hole in the limit of vanishing amplitude.
Previous work suggested the
existence of islands of stability around quasi-periodic solutions, and we 
use this new
code to examine the stability properties of approximately quasi-periodic solutions which balance energy transfer to higher modes with energy transfer to lower modes.  The evolutions provide some evidence, though not conclusively,
 for stability  of initial data sufficiently close to 
quasiperiodic solutions.

\end{abstract}

\maketitle

\tableofcontents

\section{Introduction}\label{introduction}

Numerical studies of the collapse of a scalar field in spherically symmetric spacetimes 
that approach anti-de Sitter spacetime at infinity have attracted much interest and have spawned
a number of interesting questions~\cite{Bizon:2011gg,Jalmuzna:2011qw}.
In particular, those studies showed that, beginning with some small initial pulse of scalar field, generically 
the scalar field will reflect off the boundary at infinity within finite time because of the unique causal
structure of asymptotically anti-de Sitter~(aADS) spacetimes. Importantly, these reflections will continue
after the pulse implodes through the center and makes its way to the boundary again. This process of reflection
repeats
until the pulse is sufficiently sharpened
to form a black hole. Indeed, their work suggests that for any initial amplitude, $\epsilon$, 
a weakly turbulent instability
results in sharpening sufficient to produce gravitational collapse within a time $t_\mathrm{toc}$
that scales as $1/\epsilon^2$.

The suggestion of Refs.~\cite{Bizon:2011gg,Jalmuzna:2011qw} was that the eventual collapse of even the
smallest amplitude $\epsilon$ indicated that AdS was itself unstable to the introduction of
any scalar energy, consistent with previous mathematical expectations~\cite{dafermos_holzegel,Anderson:2006ax}. However, numerical evolutions
face, at the very least, two significant limitations in this regard. The first is that any such
evolution must consider a finite amplitude $\epsilon$, and so even if that amplitude collapses,
the result does not indicate what happens for smaller amplitude. The second, just as important
as the first, is that any evolution extends to only finite time, and so even if an evolution does
not collapse for some time, it does not indicate whether collapse occurs for a time after one has stopped
the evolution.

In light of these difficulties in numerical simulation, 
one can instead consider the problem as a perturbation problem about pure AdS. 
Refs.~\cite{Bizon:2011gg,Jalmuzna:2011qw} carried out such a perturbation study and found a resonant instability
in which the normal modes couple to excite other, higher modes of the system. 
These excitations of higher modes involve the transfer of energy to
shorter wavelengths and the instability is therefore called 
weakly turbulent. 

Given the evidence of an instability from perturbation theory for 
infinitesimal $\epsilon$ and the numerical evidence for collapse at finite
$\epsilon$, it made sense to assume that the two are connected and that black 
hole formation ensues for any value of $\epsilon$.
However, further numerical
studies found that not all configurations of scalar pulses appear to collapse~\cite{Buchel:2013uba}. In particular,
widely distributed data, for sufficiently small initial amplitude, demonstrated times of collapse that become
very large quickly, and that were therefore suspected to avoid collapse. 

Recent work suggests that at least some of the configurations  that avoid collapse on long time scales are located in
a type of phase space within islands
of stability anchored by quasi-periodic~(QP) solutions~\cite{Green:2015dsa}. Such QP solutions balance transfer to higher modes
with transfer to lower modes so that their energy spectrum is static. Perturbation analysis suggests that these solutions are
stable. 

Here, we use a newly constructed open source code to study the numerical evolution of such solutions, and these evolutions appear consistent with stability~\cite{openadscl}.
%
%
%
%
This code serves as a testbed for using OpenCL, a recently developed open standard to allow for the use of
heterogeneous computing.


\section{Numerical System}\label{system}
Consider a spherically symmetric spacetime that is asymptotically
AdS adopting the metric
\begin{equation}
ds^2=\frac{1}{\cos^2 x}\left(-A e^{-2\delta} dt^2+A^{-1} dx^2+\sin^2 x\ d\Omega^2\right),
\label{metric}
\end{equation}
where $\delta(x,t)$  and $A(x,t)$ are
metric functions depending on time $t \in
(-\infty, \infty)$ and the radial coordinate $x\in [0,\pi /2]$.
We introduce a scalar field, $\phi(x,t)$, and adopt auxiliary
variables to cast its evolution in first order form with
$\Pi\equiv e^\delta\dot\phi/A$ and $\Phi\equiv \phi'$
where a dot and prime denote partial derivatives with respect to $t$ and $x$, respectively.
The equations of motion then include the following evolution equations
\begin{eqnarray}
\dot \Phi & = & \left( A e^{-\delta} \Pi \right)' \label{eq:phi}\\
\dot \Pi  & = & \frac{1}{\tan^2 x} \left( \tan^2 x A e^{-\delta} \Phi \right)' \label{eq:pi}\\
\dot A    & = & -2 \sin x \cos x A^2 e^{-\delta} \Phi \Pi \label{eq:adot}.
\end{eqnarray}
The system also includes two constraint equations
\begin{eqnarray}
A' & = & \frac{1+ 2 \sin^2 x}{\sin x \cos x} \left( 1- A \right)
         - \sin x \cos x A \left( \Phi^2 + \Pi^2 \right) \label{eq:aprime}\\
\delta' & = & - \sin x \cos x \left( \Phi^2 + \Pi^2 \right). \label{eq:delta}
\end{eqnarray}
More detail about the system can be found in Ref.~\cite{Maliborski:2013via}.

The code uses a Runge-Kutta method of lines to evolve Eqs.~(\ref{eq:phi}-\ref{eq:adot})
using fourth order accurate finite differences to approximate the spatial derivatives.
The code includes second, third, and fourth order accurate Runge-Kutta and the results presented
here use the third-order accurate option. 

These fields are subject to boundary conditions at the extremes of the computational domain. At $x=0$,
the conditions enforce regularity and local flatness
\begin{eqnarray}
\Phi(0,t) & = & 0\\
A(0,t) & = & 1.
\end{eqnarray}
At the outer boundary, $x=\pi/2$, we enforce that no energy is incoming via
\begin{eqnarray}
\Phi(\pi/2,t) & = & 0\\
\Pi(\pi/2,t) & = & 0\\
A(\pi/2,t) & = & 1.
\end{eqnarray}
The metric field $\delta(x,t)$ determines the time coordinate and allows for some freedom. We make
the choice that $t$ measures proper time at the outer boundary by enforcing
\begin{equation}
\delta(\pi/2,t)  =  0.
\end{equation}
This choice is called the ``boundary time'' gauge as discussed in Ref.~\cite{Craps:2014vaa}.

These operations (computing spatial derivatives and stepping in time) can be spread 
across the many threads of a GPU device necessary to achieve good parallel performance.
However, solving the spatial constraint Eqs.~(\ref{eq:aprime}) and~(\ref{eq:delta}) generally
involves stepping from one boundary to the other, integrating along. Because this process is
order dependent, it cannot naively be distributed among many threads. 

We therefore evolve the metric function $A(x,t)$ at every step instead of integrating
Eq.~(\ref{eq:aprime}). However, we find that periodically integrating $A(x,t)$ over space
helps improve stability and accuracy. For $\delta(x,t)$ there
is no choice other than integrating Eq.~(\ref{eq:delta}) in space at every
time step. We therefore transfer information for the current time-step to the CPU host, 
integrate the equation there, and transfer the solution for $\delta$ back to the GPU device 
(see Section~\ref{sec:implementation} for more details about the details of the implementation).

\begin{figure}[h]
\centering
\includegraphics[width=9.0cm,angle=0]{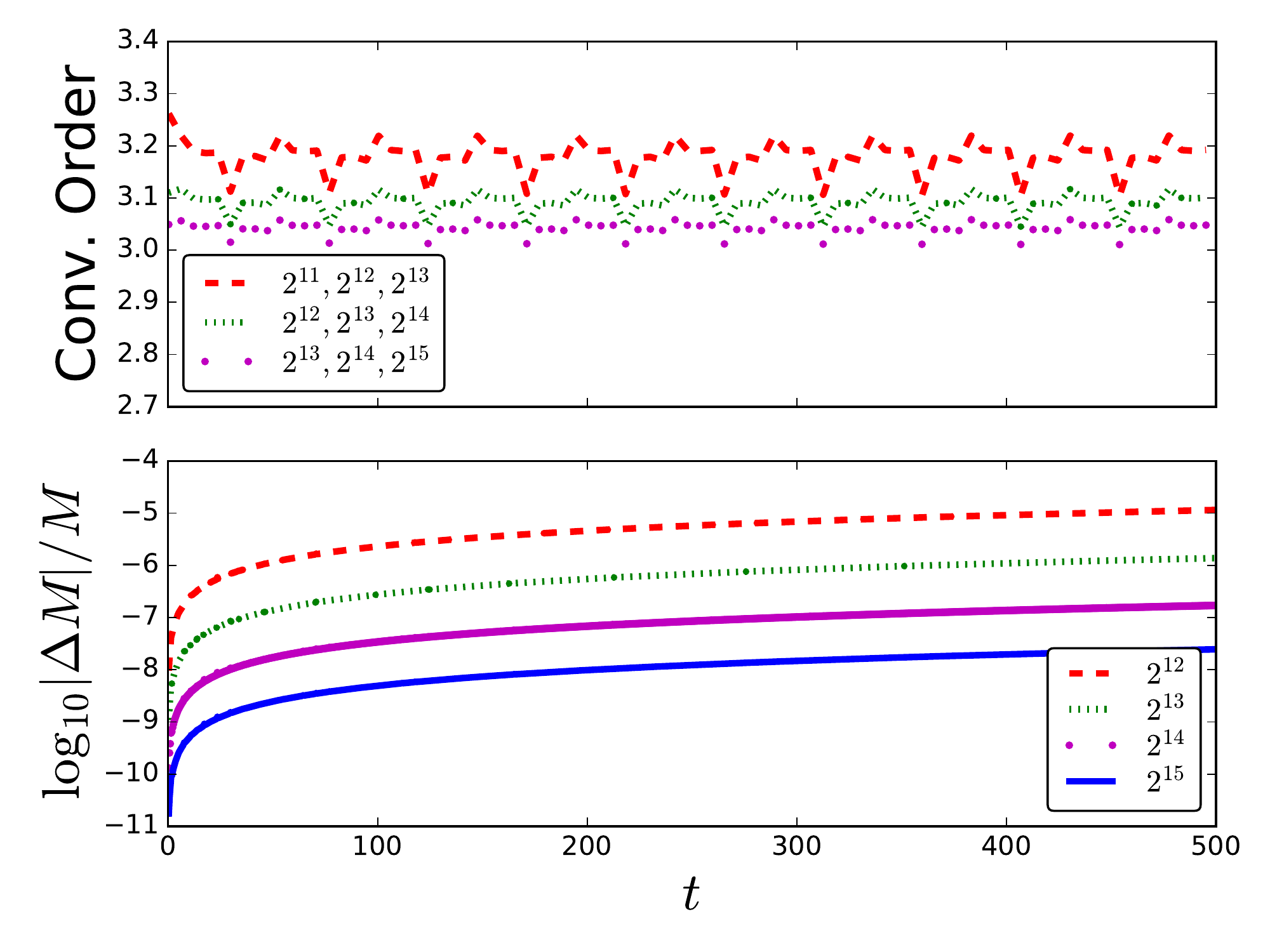}
\caption{ \textit{Demonstration of high order convergence.} 
\textbf{(Top:)}~The order of convergence for a series of increasing resolutions obtained for the field $\Pi$.
The code converges to third order in the grid spacing.
\textbf{(Bottom:)}~The numerical mass loss as a function of time for each resolution is shown. That the mass loss is small and converges to zero indicates consistency with the continuum equations.
}
\label{fig:convergence}
\end{figure}

To test the fidelity of our numerical results to continuum solutions of the system,
we monitor and test a number of aspects. We check that solutions converge to a unique
solution as the resolution is increased. We check that the solutions are consistent
with the system by monitoring the total mass of the spacetime and the residuals of
the constraint equations. In particular the mass is found by integrating
$M = \int_0^{\pi/2} \tan^2 x A \left( \Phi^2 + \Pi^2 \right) dx$.
We demonstrate both convergence and consistency of the code for one
example in Fig.~\ref{fig:convergence}. We have also compared results of this code with
another code as used in Ref.~\cite{Buchel:2013uba}.

Black hole formation is signalled by the decrease of $A(r,t)$ to zero. Because
these coordinates become singular in the presence of a horizon, the code
stops when the minimum value of $A$ becomes sufficiently small. A reasonable
value for the time of collapse is achieved by recording the coordinate
time when this minimum reaches the arbitrary, but small, value of $0.01$.

At any given time, the current numerical solution can be decomposed into a spectrum
of normal modes. This decomposition makes contact with perturbation theory, and, in particular,
provides the three quantities that are conserved within the TTF formalism~\cite{Craps:2014jwa,Buchel:2014xwa}.

We can use these normal modes to define initial data. To begin an evolution,
one must define $\Phi(x,0)$ and $\Pi(x,0)$ and solve Eq.~(\ref{eq:aprime}) for $A(x,0)$ and Eq.~(\ref{eq:delta}) for $\delta(x,0)$. 
Given some spectrum, $E(k)$, we
can then define the initial scalar field with that decomposition.
Because our intent is to study approximate QP solutions, these are the only initial data
we consider here.

\begin{figure}[h]
\centering
\includegraphics[width=9.0cm,angle=0]{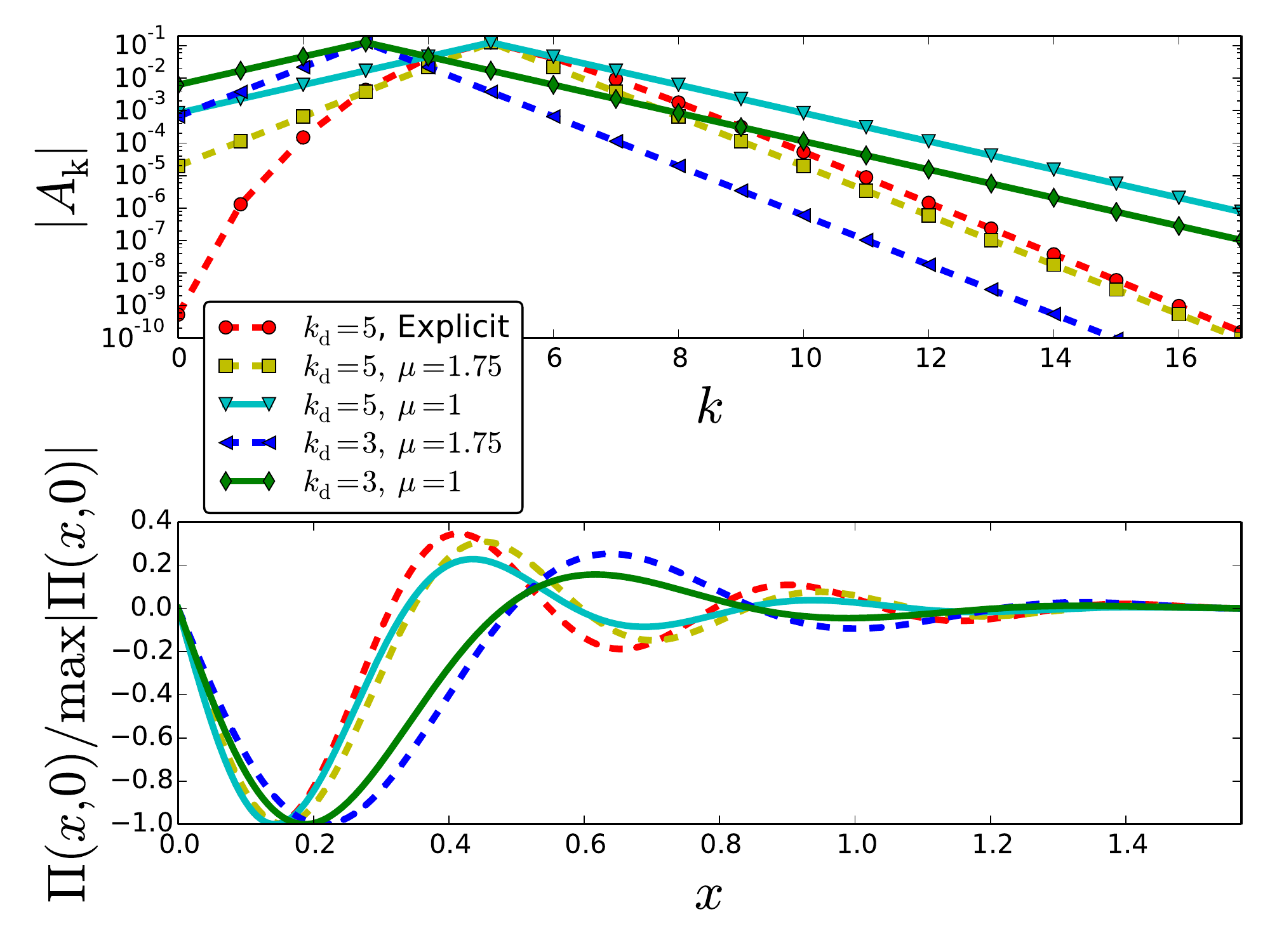}
\caption{ \textit{Initial data:}
\textbf{(Top:)} The mode amplitude spectrum for the initial data studied here.
A solution which explicitly solves the conditions of quasi-periodicity~(QP) is shown
along with approximate QP solutions defined by Eq.~\ref{eq:approxQP}.
\textbf{(Bottom:)} The spatial profile of the initial data (in which $\Phi(x,0)=0$).
}
\label{fig:ID}
\end{figure}

In what follows, we consider the evolution of one particular QP explicitly constructed in~\cite{Green:2015dsa}
as well as approximate QPs described with mode amplitudes, $A_k$, which fall-off exponentially from
a dominant mode $k_\mathrm{d}$. We parameterize these approximate solutions in terms of 
approximately quasiperiodic solutions as
\begin{equation}
A_k = {\epsilon}  e^{-\mu |k-k_\mathrm{d}|},
\label{eq:approxQP}
\end{equation}
where $k$ varies over the various modes, $\mu$ is the rate of fall-off about the dominant mode, $k_\mathrm{d}$.
The parameter $\epsilon$ represents an overall amplitude.
We show both the amplitude spectrum and initial scalar profile for some of these initial data in Fig.~\ref{fig:ID}.

The temperature is independent of $\epsilon$ and characterizes the distribution
of energy among the modes~\cite{Green:2015dsa}.
The temperature of the explicit solution for $k_\mathrm{d}=5$ is $T=13.1$ whereas the approximate QP solutions
for the same $k_\mathrm{d}=5$
have temperatures of $13.0$ and $13.1$ for $\mu=1.75$ and $\mu=1$, respectively.
The value $\mu=1.75$ is chosen because its spectrum for $k_\mathrm{d}=5$ roughly
matches the explicit solution.
The approximate QP solutions for the $k_\mathrm{d}=3$
have temperatures of $9.03$ and $9.16$ for $\mu=1.75$ and $\mu=1$, respectively.

\begin{figure}[h]
\centering
\includegraphics[width=9.0cm,angle=0]{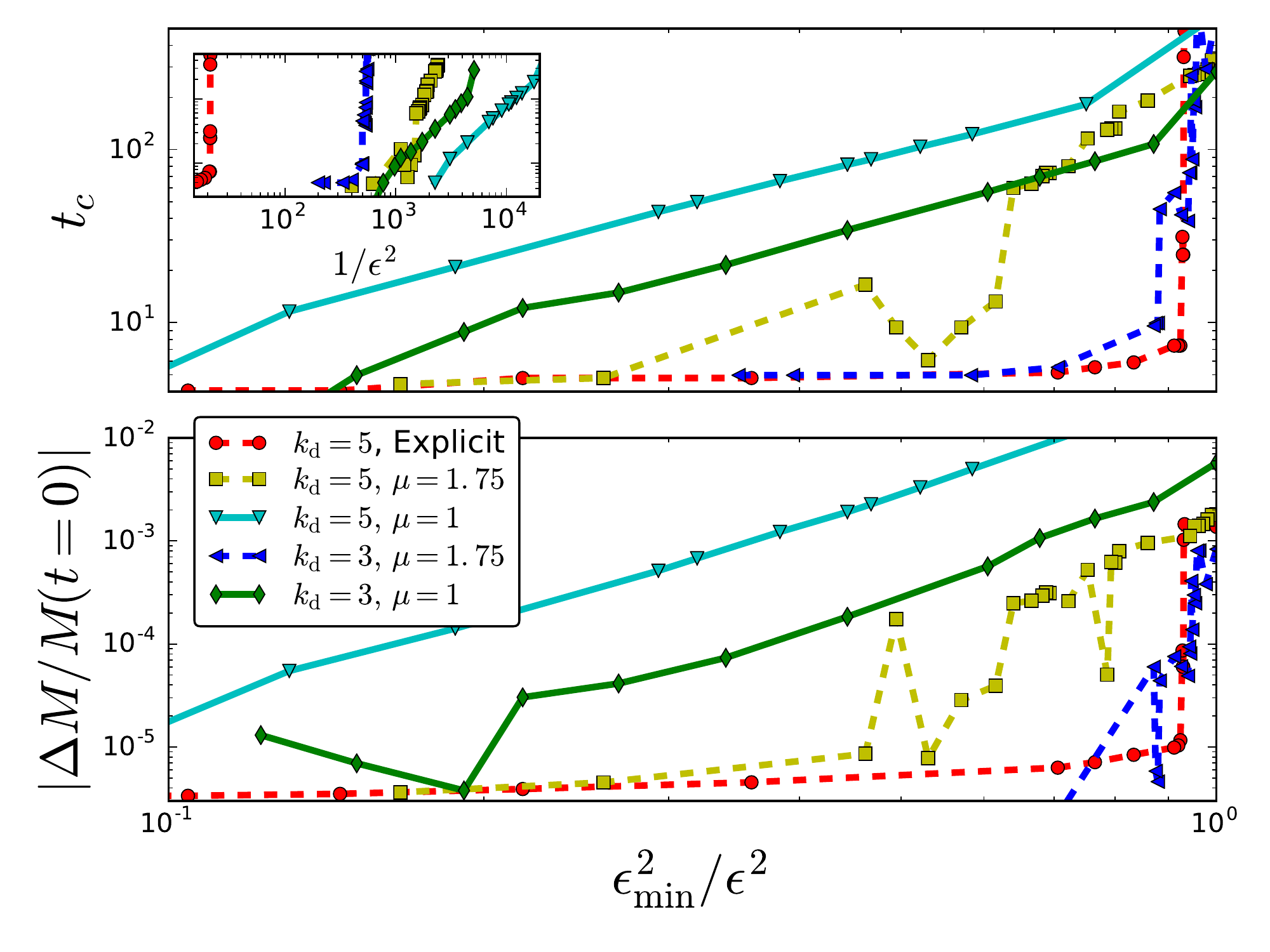}
\caption{ \textit{Time of collapse of the QP families studied here.}
To accommodate all the families on the same plot, the horizontal axis is
rescaled in terms of the smallest amplitude plotted for that family, $\epsilon_\mathrm{min}$.
\textbf{(Top:)} The times of collapse for the families studied here. 
The inset shows these same times without the rescaling on a log-log plot.
Families with $\mu=1$  demonstrate a time of
collapse that scales with $1/\epsilon^2$ consistent with instability.
In contrast, the explicit solution and those with $\mu=1.75$ deviate
significantly from such a scaling.
\textbf{(Bottom:)} The mass loss associated with each run. These runs generally 
use $2^{15}$ points which helps keep the mass loss to less than a percent.
}
\label{fig:toc}
\end{figure}

\section{Results}\label{results}

In this work, we are primarily concerned with establishing whether given
families of initial data given by Eq.~(\ref{eq:approxQP}) and parameterized by $\mu$ and $k_d$
follow the typical behavior in which the time of collapse, $t_c$,
scales with $1/\epsilon^2$ as predicted by the turbulent instability.
Because of the computational challenges in determining whether a given family is indeed
immune to the instability (for a discussion of some of these difficulties,
see Ref.~\cite{Balasubramanian:2015uua}), we instead compute $t_c$ for a series of
decreasing amplitudes, $\epsilon$, and look for deviations from this $t_c \propto 1/\epsilon^2$
scaling. If significant deviations appear, we take this behavior as suggestive, but not
conclusive, evidence for stability.
For an example of this, one can look at Figs.~5-9 of Ref.~\cite{Buchel:2013uba}.
Regardless of whether indeed such a deviation indicates stability, the deviation represents a
qualitative change and thus may help guide analytic treatments of this instability.

In particular, we evolve for a maximum time, $t_\mathrm{max}=500$, monitoring the evolution for collapse.
This choice is mostly arbitrary, balancing computational cost with the desire to be sufficiently long
as to capture the behavior of the instability.
We likewise monitor the mass loss ensuring that it remains small. In Fig.~\ref{fig:toc} we plot both
the time of collapse and the mass loss for the same families presented in Fig.~\ref{fig:ID}.
These quantities are plotted versus $1/\epsilon^2$ rescaled by the minimum $\epsilon$ value for
that family in order to display all families on the same two plots. 
The inset of the plot shows the families without this rescaling.


The top panel of Fig.~\ref{fig:toc} shows two contrasting phenomena. For some families,
$t_c$ is indeed linearly related to $1/\epsilon^2$ whereas the other families deviate from
such linearity. More particularly, families with $\mu=1$ demonstrate collapse times
characteristic of the instability, whereas the explicit QP solution along with families with
$\mu=1.75$ demonstrate different behavior.

Note that the $k_d=5$, $\mu=1.75$ family appears quite similar to the explicit QP solution in
Fig.~\ref{fig:ID}. We take this similarity to indicate that the $\mu=1.75$ families are generally
closer to their respective QP solutions. These numerical results therefore suggest that the
deviations seen for such families is consistent with the QP solutions having some radius of stability
and that the $\mu=1$ families are outside this radius of the QP solutions.

%
%
%
%

\section{Computational Aspects}\label{computational}

\subsection{What is OpenCL?}
Accelerator devices (such as GPUs, FPGAs, DSPs, and the Xeon Phi, etc.) are gaining tremendous momentum in the HPC world today. In fact, accelerators are ubiquitous in the very top supercomputers today\footnote{http://top500.org}, and their use is largely what has allowed the performance of these machines to enter into the peta-scale regime and beyond. However, different accelerators traditionally employed a rather unfamiliar and specialized programming model that often required advanced knowledge of their hardware design. In addition, they historically have had their own vendor- and design- specific software development kits~(SDK) with little in common amongst them.

However, in 2010, an open standard was proposed by Apple Inc. to ``unify'' the software development for all these different computer architectures under a single parallel programming standard -- the {\em Open Computing Language} (OpenCL)\footnote{https://www.khronos.org/opencl/}. OpenCL is a free, open standard that is vendor and platform neutral. It allows one to write high-performing, yet, {\em portable} parallel code that executes on a wide variety of processor architectures including CPUs and all accelerator devices. The major processor vendors (Nvidia, AMD/ATI, IBM, Intel, Altera, etc.) have adopted the OpenCL standard and have released support for it on their hardware. 

In brief, OpenCL incorporates the changes necessary to the programming language C that allow for parallel computing on all these different architectures. In addition, it establishes numerical precision requirements to provide mathematical consistency across the different hardware and vendors. In OpenCL, the accelerator (called {\em device}) is accessible to the CPU (called {\em host}) as a co-processor with its own memory. The device executes a function (usually referred to as a {\em kernel}) in a data-parallel model, which means that a number of threads run the same program on different data. Computational scientists must rewrite the performance intensive routines in their codes as OpenCL kernels that would be executed on the compute hardware. 

The OpenCL API provides the programmer various functions from locating the OpenCL enabled hardware on a system to compiling, submitting, queuing, synchronizing the compute kernels and performing memory management on the hardware. Finally, it is the OpenCL runtime that actually executes the kernels and manages the needed data transfers in an efficient manner. Computation is performed within groups of threads that are assembled into a {\em workgroup}. If the accelerator has a large number of compute-units, it can process them in parallel. That architecture allows one to farm out a large number of calculations into a series of groups and execute them in parallel. The threads then scatter, doing all of their farmed computations on the different compute-units, and synchronize at the completion of their assigned tasks.

\subsection{OpenCL Implementation}
\label{sec:implementation}
A data-parallel model is relatively straightforward to implement in a standard finite-difference code like ours. We simply perform a domain decomposition of our computational grid and allocate the different parts of the grid to different compute-cores. More specifically, on a many-core GPU accelerator device, each thread computes the entire evolution of the system for a single grid-point (i.e. value of $x$). To access data from neighboring grid-points, we make extensive use of the ``shared'' nature of the GPU device's {\em global memory} thus  highly simplifying the communication between the GPU compute-cores. We make this simplification for the stated goal of keeping the code's portability intact, even if it impacts performance to some extent. 

The entire numerical scheme is split across 15 separate kernels that are queued to execute in succession at each time-step. As mentioned in the previous section, the primary scheme in use by the code is a relatively straightforward method-of-lines scheme with fourth-order finite differencing and Runge-Kutta time-stepping. The code is freely available under an open source license via a public {\tt bitbucket} repository~\cite{openadscl}. 

It is worthwhile to make two remarks regarding our data-parallel OpenCL implementation. We choose to execute nearly all portions of the numerical scheme, even the ones that have a relatively low {\em arithmetic intensity} on the GPU device, to help minimize data movement back to the CPU host system's main memory. This data movement occurs over the slow PCI bus, and therefore must be minimized for good performance. A significant challenge is posed by the constraints Eqs.~\ref{eq:aprime} and~\ref{eq:delta} on the physical variables $\delta$ and metric $A$. In our numerical scheme, these are to be solved via a simple integration or sum over the entire computational grid. Due to the inherently serial nature of those types of computations, it is difficult to execute them on a many-core GPU device efficiently. Different approaches were tested, from using a single GPU device core as essentially a serial processor to perform the integrals, to solving those equations using a relaxation method. Interestingly, we found that the most effective approach was simply to copy the $\delta$ and $A$ data arrays back to main memory, and process them there using the system's CPU cores instead (and copy them back to the GPU device).  However, we suspect more effort at a parallel
computation would likely perform better.

\subsection{Performance Results}
In this section, the overall computational performance on recent hardware is documented in detail. The host system is an Ubuntu 14.04 Linux server with an 8-core, 4.7 GHz AMD FX-9590 (liquid-cooled) processor. The system main memory is 24 GB and has a fast 120 GB SSD disk. The accelerator devices we tested are the recent generation Nvidia Tesla series CUDA GPU K40 and the AMD Radeon HD R9-295x2 GPU. Details on the specifications of these devices are provided below in Tables~\ref{comp} \& \ref{comp2}. 

The performance gains derived from GPU-acceleration are presented in the Tables~\ref{comp} \& \ref{comp2} in the form of an overall speedup over the CPU-only case. For the cases in which the physical variables $A$ and $\delta$ are not dynamical i.e. Table~\ref{comp2}, we obtain {\em order-of-magnitude} gains in the overall performance of AdSCL. This is a fairly typical outcome that has been appreciated by the computational science community for several years now. For the cases in which $A$ and $\delta$ are evolved i.e. Table~\ref{comp}, the speedup is substantially less due to the extensive communication needed between the GPU and CPU cores (as remarked in the previous section). 
{However, there is still a solid benefit from the GPU-acceleration even in that
case, and allows one to explore regimes that may not have been practical before.}

\begin{table}
  \centering
     \begin{tabular}{|c|c|c|c|c|} \hline
   {\bf Name} & {\bf Type} & {\bf GHz} & {\bf Cores} & {\bf Speedup}\cr
    \hline \hline
    {\bf AMD FX-9590}   & CPU &   4.7    &  8    & {\bf 1x} \cr \hline
    {\bf Nvidia Tesla K40}  & GPU &   0.8    &  2880   & {\bf 3.6x} \cr \hline
    {\bf AMD Radeon R9} & GPU &   1.0    &  2816  & {\bf 3.0x} \cr \hline
   \end{tabular}
 \caption{{\footnotesize This table depicts the relative values for overall performance for several variants of recent generation GPUs. The baseline system here is an AMD FX 8-core, 4.7 GHz CPU running our AdSCL OpenCL Code with $N=32K$. Full {\em double-precision} floating point accuracy was used for these tests. }}
\label{comp}
\end{table}

\begin{table}
  \centering
     \begin{tabular}{|c|c|c|c|c|} \hline
   {\bf Name} & {\bf Type} & {\bf GHz} & {\bf Cores} & {\bf Speedup}\cr
    \hline \hline
    {\bf AMD FX-9590}   & CPU &   4.7    &  8    & {\bf 1x} \cr \hline
    {\bf Nvidia Tesla K40}  & GPU &   0.8    &  2880   & {\bf 36x} \cr \hline
    {\bf AMD Radeon R9} & GPU &   1.0    &  2816  & {\bf 29x} \cr \hline
   \end{tabular}
 \caption{{\footnotesize This table depicts the same quantities as in Table~\ref{comp}, but for the case wherein both $A$ and $\delta$ are not dynamical i.e. held fixed at $1$ and $0$ respectively, and $N=64K$. }}
\label{comp2}
\end{table}

Clearly, the key to obtaining higher performance in the dynamical $A$ and $\delta$ cases is to find a way to minimize data movement between the CPU and GPU cores. One way to do this is to use very recent processor architectures such as {\em fused} processors that contain both CPU and GPU cores on the same silicon chip. Examples of such architectures are AMD's  ``Accelerator Processing Units''~(APUs) and other ``System On a Chip''~(SoCs) such as those found in mobile devices. In such architectures, since both types of cores may access the same memory banks, one simply has to pass a pointer between the CPU and GPU cores and no data copying back-and-forth is necessary. This functionality very recently became available in OpenCL 2.0 and will be explored in detail in the future. 

%
%
\vspace{0.5cm}

\begin{acknowledgments}
It is a pleasure to thank Alex Buchel, Stephen Green, and Luis Lehner
for helpful discussions.
This work was supported in part by the NSF under grants PHY-1308621~(LIU), PHY-1607291~(LIU),
PHY-1606333~(UMassD), and PHY-1414440~(UMassD). Support was also provided by  U.S. Air Force agreement No. 10-RI-CRADA-09, 
NASA's ATP program through grant NNX13AH01G, and some computations were performed at XSEDE.
\end{acknowledgments}

\bibliographystyle{utphys}
\bibliography{paper}

\end{document}

%% file: aas_macros.tex
\def\jnl@style{\it}
\def\aaref@jnl#1{{\jnl@style#1}}

\def\aaref@jnl#1{{\jnl@style#1}}

\def\aj{\aaref@jnl{AJ}}                   
\def\araa{\aaref@jnl{ARA\&A}}             
\def\apj{\aaref@jnl{ApJ}}                 
\def\apjl{\aaref@jnl{ApJ}}                
\def\apjs{\aaref@jnl{ApJS}}               
\def\ao{\aaref@jnl{Appl.~Opt.}}           
\def\apss{\aaref@jnl{Ap\&SS}}             
\def\aap{\aaref@jnl{A\&A}}                
\def\aapr{\aaref@jnl{A\&A~Rev.}}          
\def\aaps{\aaref@jnl{A\&AS}}              
\def\azh{\aaref@jnl{AZh}}                 
\def\baas{\aaref@jnl{BAAS}}               
\def\jrasc{\aaref@jnl{JRASC}}             
\def\memras{\aaref@jnl{MmRAS}}            
\def\mnras{\aaref@jnl{MNRAS}}             
\def\pra{\aaref@jnl{Phys.~Rev.~A}}        
\def\prb{\aaref@jnl{Phys.~Rev.~B}}        
\def\prc{\aaref@jnl{Phys.~Rev.~C}}        
\def\prd{\aaref@jnl{Phys.~Rev.~D}}        
\def\pre{\aaref@jnl{Phys.~Rev.~E}}        
\def\prl{\aaref@jnl{Phys.~Rev.~Lett.}}    
\def\pasp{\aaref@jnl{PASP}}               
\def\pasj{\aaref@jnl{PASJ}}               
\def\qjras{\aaref@jnl{QJRAS}}             
\def\skytel{\aaref@jnl{S\&T}}             
\def\solphys{\aaref@jnl{Sol.~Phys.}}      
\def\sovast{\aaref@jnl{Soviet~Ast.}}      
\def\ssr{\aaref@jnl{Space~Sci.~Rev.}}     
\def\zap{\aaref@jnl{ZAp}}                 
\def\nat{\aaref@jnl{Nature}}              
\def\iaucirc{\aaref@jnl{IAU~Circ.}}       
\def\aplett{\aaref@jnl{Astrophys.~Lett.}} 
\def\apspr{\aaref@jnl{Astrophys.~Space~Phys.~Res.}}
\def\bain{\aaref@jnl{Bull.~Astron.~Inst.~Netherlands}} 
\def\fcp{\aaref@jnl{Fund.~Cosmic~Phys.}}  
\def\gca{\aaref@jnl{Geochim.~Cosmochim.~Acta}}   
\def\grl{\aaref@jnl{Geophys.~Res.~Lett.}} 
\def\jcp{\aaref@jnl{J.~Chem.~Phys.}}      
\def\jgr{\aaref@jnl{J.~Geophys.~Res.}}    
\def\jqsrt{\aaref@jnl{J.~Quant.~Spec.~Radiat.~Transf.}}
\def\memsai{\aaref@jnl{Mem.~Soc.~Astron.~Italiana}}
\def\nphysa{\aaref@jnl{Nucl.~Phys.~A}}   
\def\physrep{\aaref@jnl{Phys.~Rep.}}   
\def\physscr{\aaref@jnl{Phys.~Scr}}   
\def\planss{\aaref@jnl{Planet.~Space~Sci.}}   
\def\procspie{\aaref@jnl{Proc.~SPIE}}   